\documentclass[aps,amsmath,reprint]{revtex4-2}
\usepackage{graphicx}
\usepackage{dcolumn}
\usepackage{bm}
\usepackage{amsfonts}
\usepackage{array}
\usepackage[colorlinks,linkcolor=blue,anchorcolor=blue,citecolor=blue]{hyperref}

\begin{document}

\title{Realizing downstream access network using continuous-variable quantum key distribution}
\author{Yundi Huang$^{1}$}
\author{Tao Shen$^1$}
\author{Xiangyu Wang$^{1}$}
\thanks{xywang@bupt.edu.cn}
\author{Ziyang Chen$^2$}
\thanks{chenziyang@pku.edu.cn}
\author{Bingjie Xu$^3$}
\author{Song Yu$^{1}$}
\author{Hong Guo${^2}$}

\affiliation{$^1$State Key Laboratory of Information Photonics and Optical Communications, Beijing University of Posts and Telecommunications, Beijing 100876, China}

\affiliation{$^2$State Key Laboratory of Advanced Optical Communication Systems and Networks, Department of Electronics, and Center for Quantum Information Technology, Peking University, Beijing 100871, China}

\affiliation{$^3$Science and Technology on Communication Security Laboratory, Institute of Southwestern Communication, Chengdu 610041, China}
\date{\today}

\begin{abstract}
Quantum key distribution (QKD) which enables the secure distribution of symmetric keys between two legitimate parties is of great importance in future network security.
Access network that connects multiple end-users with one network backbone can be combined with QKD to build security for end-users in a scalable and cost-effective way.
Though previous QKD access networks are all implemented in the upstream direction, in this paper, we prove that downstream access network can also be constructed by using continuous-variable (CV) QKD.
The security of the CV-QKD downstream access network is analyzed in detail, where we show the security analysis is secure against other parties in the network.
The security analysis we proved corresponds to the downstream access network where only passive beamsplitter is sufficient to distribute the quantum signals and no other active controls are demanded. Moreover, standard CV-QKD systems can be directly fitted in the downstream access network, which makes it more applicable for practical implementations.
Numerous simulation results are provided to demonstrate the performance of the CV-QKD downstream access network, where up to 64 end-users are shown to be feasible to access the network.
Our work provides the security analysis framework for realizing QKD in the downstream access network which will boost the diversity for constructing practical QKD networks.

\end{abstract}

\maketitle


\section{Introduction}
Quantum key distribution (QKD) which is designed for addressing the vulnerability of the classical data encryption schemes against malicious quantum computers may provide an ultimate solution for future network security~\cite{RevModPhys.74.145, Arxiv.1906.01645v1, RevModPhys.81.1301}.
Continuous-variable (CV) QKD as a branch of QKD protocols can be implemented with only the cost-effective and off-the-shelf components~\cite{PhysRevLett.88.057902, RevModPhys.84.621, diamanti2015distributing, RevModPhys.92.025002}. The coherent state and homodyne detection CV-QKD protocol that reveals great simplicity and availability has thus attracted much efforts in its experimental realization~\cite{PhysRevA.76.042305, PhysRevA.76.052323, jouguet2013experimental, huang2016long, PhysRevApplied.10.064028, guo_2021_FundRes}. Recently, the long-distance of CV-QKD experimental implementation has achieved a transmission distance of over 202 km in ultralow loss fiber~\cite{PhysRevLett.125.010502}, and the practical field test has shown the feasibility of CV-QKD in the metropolitan distance~\cite{zhang2019continuous}. Theoretical analysis has also proved the upper bounds of the secret key rate of the CV-QKD protocols~\cite{PLOB, PLOB2018}.
The rapid developments in both the theoretical analysis and the experimental implementations have led to more attentions in extending the standard point-to-point QKD to the more practical environment where QKD is performed in networks~\cite{salvail2010security, peev2009secoqc, huang2016field, chen2021an}.
In reality, diverse network structures are demanded in order to build a globally quantum internet~\cite{lloyd2004infrastructure, kimble2008quantum, pirandola2016physics, wehner2018quantum}.

Access network that connects multiple subscribers with one network backbone is a modern network necessity. More importantly, it enables that the end-users are able to access to the network in a cost-effective and scalable way.
In a typical access network, the end-users are called the optical network units (ONUs), and the network node which located at the network backbone is the optical line terminal (OLT).
The intermediate node that connects the ONUs and the OLT are called the optical distribution network (ODN).
The ODN can combine signals from the ONUs or distribute signals from the OLT, according to the direction of signal transmissions. For upstream signals that are sent from the ONUs, the passive beamsplitter (BS) and the wavelength-division multiplexing device are usually adopted in the ODN to combine signals in the time-division multiplexing scheme. And for downstream signals, signals are generated from the OLT, and then distributed to every ONU in the network. The passive BS are mostly utilized as the passive optical network (PON) in the ODN.

For the purpose of achieving QKD between the OLT and the ONU, intuitively, both upstream transmission scheme and downstream transmission scheme seems to be possible, since signals are transmitted between the OLT and the ONUs.
Early considerations have investigated the possibility of applying QKD into the access network~\cite{a6253218, a6582861}, and the first quantum access network is experimentally demonstrated for discrete-variable (DV) QKD~\cite{frohlich2013quantum}. Later on, quantum access networks of DV-QKD with different implementations such as plug-and-play scheme or measurement-device-independent scheme are also proposed~\cite{park2020user, wei2020high}. Nevertheless, all of the previous reported demonstrations have implemented an upstream access network structure where the receiver end is placed at the network backbone OLT.

The downstream access network on the other hand may offer extra advantages. For upstream access network, signals are generally applied the time-division multiplexing scheme where signals are demanded to be precisely allocated to the corresponding time slots.
The time slot allocation is critical especially when more ONUs are connecting to the network, or when the network traffic is high, in which cases the preserved time slots would inevitably be narrowed down. Crosstalk may also arise under such circumstances which can result in the deterioration of the quantum signals.
Such limitations have very less impact on the downstream access network, since no multiplexing technique is demanded for the downstream access network. It also means no active controls are required for the ODN.
What is more, the downstream access network in the classical networks normally deploys a higher physical layer line rate than the upstream access network (downstream/ upstream: 2.488 Gbps/ 1.244 Gbps)~\cite{ITU2}, which suggests that the downstream access network can take much larger signal flows in the network.
It is thus of great importance to apply QKD in the downstream access network structure.

For the task of establishing symmetric keys in the downstream direction, the keys should be private with only the OLT and the ONU, such secrecy should be against all other parties in the network. While, in the downstream access network, the quantum signals are broadcasted to all ONUs in the network. Thus, in order to adopt QKD in the downstream access network, it is critical to demonstrate that even with quantum signals distributed to all ONUs, the final key is secure against other parties in the network.
In this paper, we prove the security of the downstream access network using CV-QKD.
The security analysis of using CV-QKD in the downstream access network is studied in detail, where we show secret key rate against other parties in the network can be extracted.
We further simplify the security analysis so that it can be conducted only by the OLT and the corresponding ONU without the complicated channel parameters calibrations and the assistance of any other parties in the network.
The corresponding practical implementation can directly utilize standard CV-QKD systems which suggests the great availability of the CV-QKD downstream access network. Numerous simulations are provided to demonstrate the performance of the downstream CV-QKD access network.

\begin{figure*}[ht]
\centerline{\includegraphics[width=15.0cm]{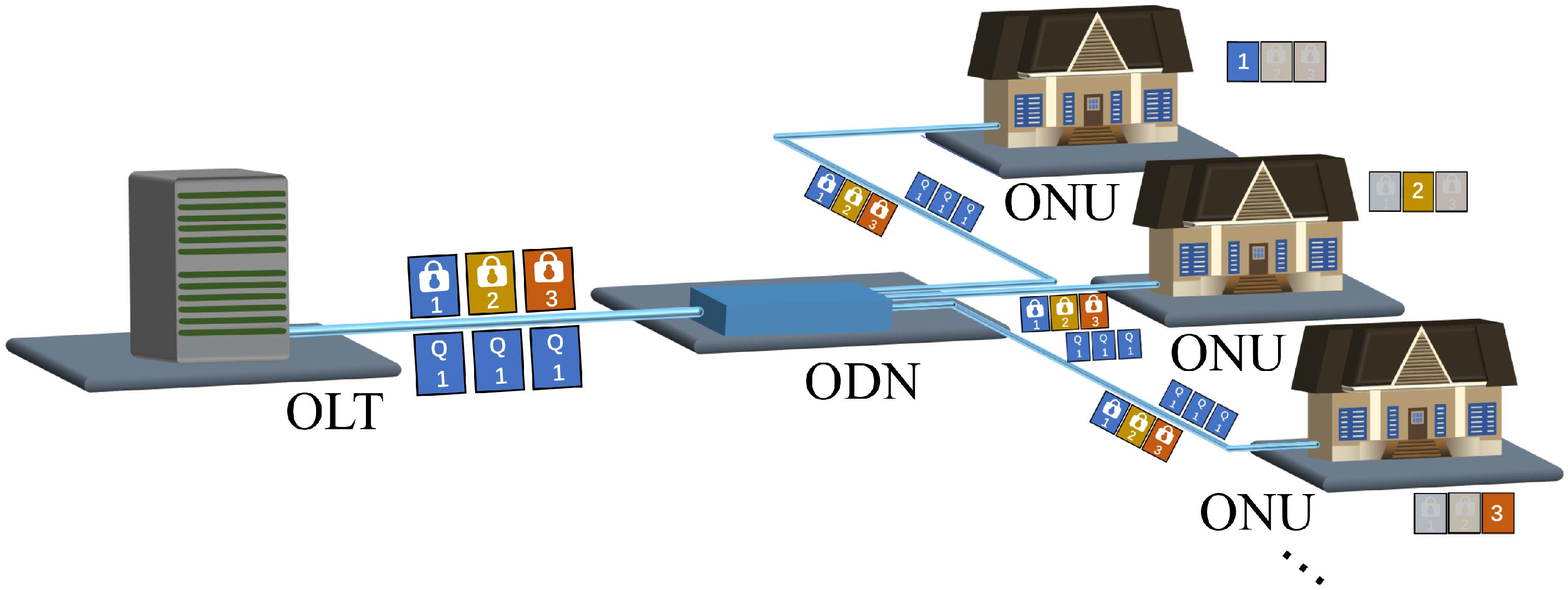}}
\caption{ (Color online) Introductory map of the downstream access network. The optical line terminal (OLT) continuously send signals to the optical distribution network (ODN), who passively separates the signals into multiple copies, and forwarded the signals to the optical network units (ONUs). Classically, data packets that to be delivered to each ONU is encrypted with keys between the OLT and the specified ONU. Although all ONUs can receive the copies of the data packets, only those packets that belongs to them can be further decrypted.
However, for the purpose of achieving QKD, it is meaningless to encrypt the key signals which is used to further construct keys. Thus, the key signals are broadcasted to each ONU, and the signals are directly accessible for all ONUs in the QKD scenario. }
\end{figure*}

The rest of the paper is organized as follows. In section~\ref{sec:2}, we quickly review the principle of classical downstream access network, then describe the CV-QKD downstream access network. Detailed security analysis is provided in section~\ref{sec:3}, and the performance of the CV-QKD downstream access network are presented in section~\ref{sec:4}. Conclusions are finally made in section~\ref{sec:5}.

\section{\label{sec:2} CV-QKD in downstream access network}

In this section, we briefly review the working principle of the downstream access network, discuss the possible implementations, and then describe the CV-QKD downstream access network.

In the classical downstream access network, the OLT continuously sent signals or data packets to the ODN. The BS which is applied at the ODN passively separate the signals, the separated signals are then transmitted to each ONU in the network through the individual fibers. Since the BS works passively at the ODN, no active selection of particular data packets is conducted, which means the data packets that are sent from the OLT are broadcasted to every ONU in the network. After receiving the signals, ONUs themselves apply the frame filter which will help them select the data packets that belongs to them, and discard other signal packets that they do not need.
Secrecy is a critical concern under such a transmission scheme, since every ONU in the network receives a copy of the data packets, it is crucial that only the specified ONU is able to know the information of that data packet, while other ONUs in the network should gain no information on the data packets that do not belong to them.
In classical networks, the secrecy is achieved by combining the data encryption technique, where the OLT encrypts the data packet before sending them to the network, as is depicted in Fig. 1. The key used in the data encryption should be private with only the OLT and the corresponding ONU so that only the corresponding ONU can decrypt the data packet and obtain the data, this maintains the secrecy since no other ONUs can decrypt the data packet without knowing the key.

Intuitively, using DV-QKD in the downstream access network may not very applicable.
The single-photon detectors which are used in the DV-QKD are more expensive for the end-users to stand. Moreover, for the downstream access network where passive BS is mostly applied in the ODN, it is impossible for the ODN to deterministically distribute particular signals to a specified ONU~\cite{frohlich2013quantum}.
Such problems on the other hand are less of difficulties for CV-QKD, the homodyne detector are much cheaper and also more robust. The quantum source used in CV-QKD like coherent states and squeezed state can deterministically distributed to the ONU. Thus, it seems that CV-QKD is a strong candidate for realizing downstream access network.

Yet, there are other concerns that need to be addressed in order to implement CV-QKD in the downstream access network.
The goal of QKD is to build symmetric secret keys only between the two legitimate parties. In the access network scenario, the key should only be accessible between the OLT and the specific ONU. While for the downstream access network, signals are broadcasted. And for QKD, it means the quantum signals transmitted in the downstream access network have no privacy, since every ONU gets a copy of them.
To show CV-QKD is available for the downstream access network, it is important to demonstrate that the final key is private against other ONUs in the network.

Next, we describe the CV-QKD downstream access network.
The practical implementation of the CV-QKD downstream access network is constructed by adopting the prepare-and-measure (PM) model.
The downstream access network for CV-QKD is essentially the same as it is for classical networks, where only passive BS is placed at the ODN, and no active optical switches or wavelength selective switches are needed.
The coherent state and homodyne detection scheme which is the most applied in the experimental demonstrations of CV-QKD is taken as an example.
In the PM scheme, Alice generates Gaussian modulated coherent states as she does in the standard CV-QKD protocols, then she sends the modulated states (quantum signals) to the network. The quantum signals  are transmitted through a single fiber until they arrived at the ODN, where the BS-PON passively splits the quantum signals into ${n}$ paths. Then the ${n}$ paths signals are transmitted to each ONU through individual fibers. ONUs can detect the received signals with the standard CV-QKD receivers. What is being different in the downstream access network is that the standard classical post-processing is conducted with one specific ONU at a time, which means that the final key is only obtained for OLT and one specific ONU.

\section{\label{sec:3} Security analysis}
In this section, we provide detailed security analysis on the CV-QKD downstream access network which is described in Section~\ref{sec:2}. The security analysis is conducted against the collective attack which is not only the most applied security analysis in practical implementations~\cite{jouguet2013experimental, huang2016long, zhang2019continuous} but also shown to be optimal under the asymptotic regime~\cite{Leverrier_PhysRevLett_2013, Leverrier_PhysRevLett_2015, Leverrier_PhysRevLett_2017}. The limitations of the standard security analysis in the downstream access network are discussed, then we prove the security analysis which is applicable to the downstream access network.

In the following, we consider a 4-ONU downstream access network scenario for instance. While in practice, more ONUs can be deployed, and the security analysis can also be directly extended.
While the practical implementations are constructed based on the PM model, its corresponding security analysis is usually conducted by using the entanglement-based (EB) model. However, the standard EB model of CV-QKD protocol is not available in the downstream access network scenario. The standard CV-QKD is designed to realize the secure distribution between two parties, while the access network will inevitably involve more parties. Also, the ODN which is located in the middle of the channel separates the entire channel into different segments.
Thus, the EB model should be modified as in Fig. 2 (a), where mode ${{B_1}}$ represents the loss introduced at the ODN, and ${{C_1}}$ to ${{C_4}}$ are the modes received by each ONUs.

\begin{figure*}[t]
\centerline{\includegraphics[width=15.0cm]{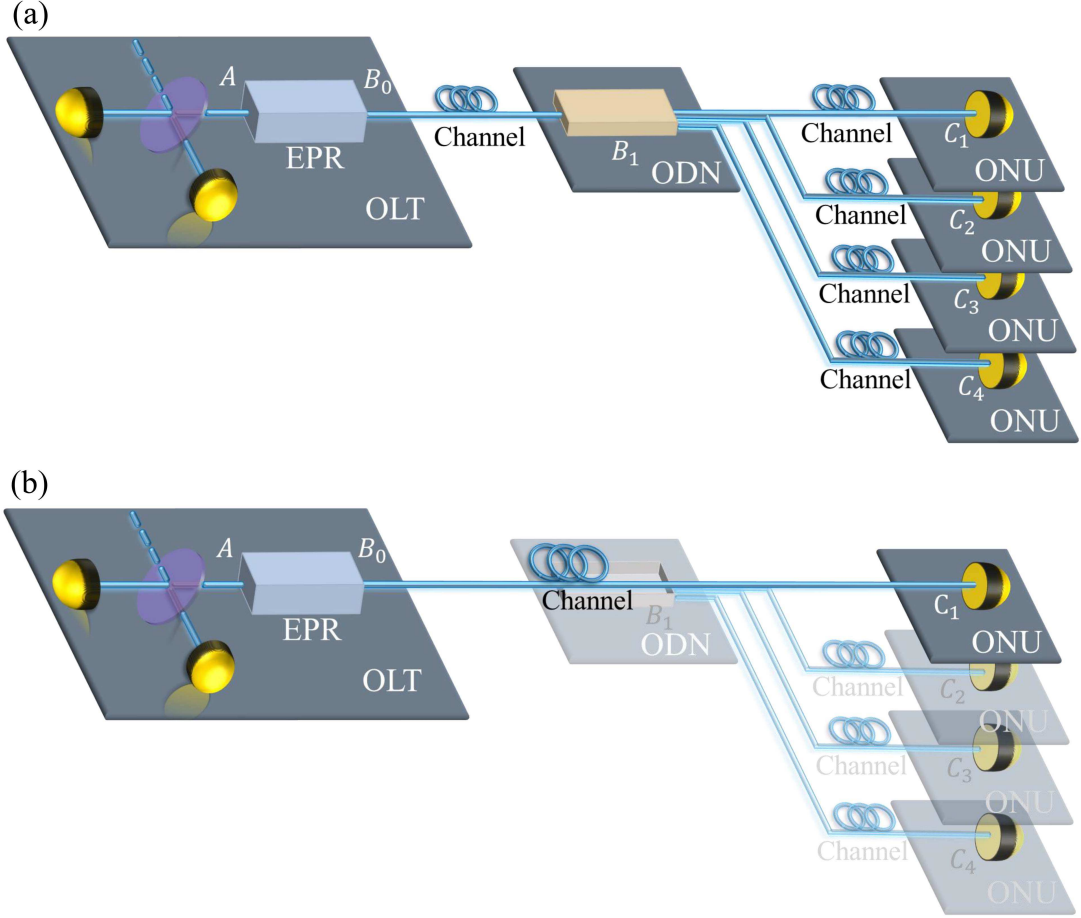}}
\caption{ (Color online) The EB model of the downstream access network where 4 ONUs are considered as an example. The model is based on the coherent state and homodyne detection scheme. Alice which is located in the OLT generates Einstein-Podolsky-Rosen (EPR) states. She then heterodyne measures one of the mode ${A}$ of the EPR state which projects the other mode ${B_0}$ into a coherent state. The state is then sent to the ODN, and the modes received by the ONUs are ${C_1}$ to ${C_4}$. (a) Extension from the standard point-to-point EB model to the downstream access network scenario. (b) The further modified EB model which is applicable for the security analysis, where only the OLT and the specified ONU are highlighted. }
\end{figure*}

The security analysis that decides the achievable secret key rate is able to subsequently be conducted based on the EB model. Suppose CV-QKD is performed with the OLT and the ONU with incoming signal mode ${C_1}$, originally, the secret key rate is calculated as~\cite{Devetak_ProcRSoc_2005} in the reverse reconciliation (RR) regime~\cite{Grosshans_Nature_2003}:
\begin{equation}
K = \beta {I_{A{C_1}}} - {\chi _{E{C_1}}},
\end{equation}
where ${\beta}$ is the reconciliation efficiency, ${{I_{A{C_1}}}}$ is the mutual information of mode ${A}$ and mode ${C_1}$.
${{\chi _{E{C_1}}}}$ represents the amount of information that Eve can obtain, which can be calculated as:

\begin{equation}
{\chi _{E{C_1}}} = S(E) - S(E|{C_1}),
\end{equation}
where ${S( \cdot )}$ stands for the Von Neumann entropy, ${S(E)}$ and ${S(E|{C_1})}$ are the entropy of Eve, and the entropy of Eve given the measurement results of mode ${C_1}$ respectively.
Standard security analysis considers Eve can purify the entire system, so that the upper bound of information that Eve can obtain can then be estimated~\cite{Garcia2006unconditional, Nava2006optimality, Leverrier2010simple, Wolf2006extremality}, more specifically, ${S(E)}$ and ${S(E|{C_1})}$ should be calculated as:
\begin{equation}
\begin{array}{l}
S(E) = S(A{B_1}{C_1}{C_2}{C_3}{C_4}),\\
S(E|{C_1}) = S(A{B_1}{C_2}{C_3}{C_4}|{C_1}).
\end{array}
\end{equation}
Thus, all modes in the right hand side of the equations need to be identified in order to decide the information that Eve can obtain, and it implies that the co-variance matrix ${{\gamma _{A{B_1}{C_1}{C_2}{C_3}{C_4}}}}$ needs to be fulfilled. This is obviously not applicable to practical realizations, because it will take too much effort to calibrate them in pracitcal systems.
It should also be noted that the security analysis should be able to be proceed without the collaboration from other ONUs in the network, since in practical the ONUs are more likely to be located in separated locations and unknown with each other.

Moreover, the security of the downstream access network should be analyzed more than the standard point-to-point CV-QKD scenario where the security is only against the eavesdropper Eve. The secrecy of the final key should also against other ONUs in the network.
Therefore, the security analysis should also subtract the amount of information that is related to other ONUs:
\begin{equation}
K' = K - {\chi _{{C_1}{C_2}}} - {\chi _{{C_1}{C_3}}} - {\chi _{{C_1}{C_4}}},
\end{equation}
where ${{\chi _{{C_1}{C_2}}}}$, ${{\chi _{{C_1}{C_3}}}}$, and ${{\chi _{{C_1}{C_4}}}}$ are the amount of correlations of other ONUs with respect to mode ${C_1}$. These terms should all be calculated to complete the security analysis. However, these terms are difficult to be determined in practical because of the complicated calibrations demanded on other ONUs in the network. Even the EB model in Fig. 2 (a) is modified to fit in the downstream access network scenario, it is still inappropriate for the security analysis.

Given the above considerations, the security analysis should be decided in a more efficient way, also, the calculated secret key rate should be secure against other ONUs in the network.
The solution is to consider other ONUs in the downstream access network as being controlled by one stronger Eve.
Subsequently, the EB model is further modified as in Fig. 2 (b). In this case, the OLT Alice still holds one mode of the EPR state, though all ONUs in the network receive the quantum signals, only the ONU which is engaged to the current round of QKD is activated, as is highlighted in Fig. 2 (b).
All other network nodes are silenced which implies these nodes are considered to be part of the untrusted environment or Eve.

It should be noted that Eve in the security analysis is defined as owing infinite computational power or having no restrictions on her manipulation techniques on the quantum signal~\cite{RevModPhys.81.1301}, which suggests that Eve is herself very strong.
In practice, though ONUs correspond to end-users, for the security analysis that are conducted between the OLT and the activated ONU, other ONUs are treated as part of Eve. In this sense, Eve is said to be ``stronger".
Such a model can help build up security agasint the vicious network nodes.
In the case of some ONUs are indeed malicious, the key information that they can obtain is still restricted by the Holevo bound, which is estimated as the upper bound of information that Eve can obtain in the standard security analysis.

The information that regards the potential malicious parties can obtain, can then be carried out by estimating the upper bound information of an strengthened Eve ${E'}$.
Thus, ${S(E)}$ and ${S(E|{C_1})}$ in eq. (3) can be rewritten as ${S(E')}$ and ${S(E'|{C_1})}$:
\begin{equation}
\begin{array}{l}
S(E') = S(E{B_1}{C_2}{C_3}{C_4}) = S(A{C_1}),\\
S(E'|{C_1}) = S(E{B_1}{C_2}{C_3}{C_4}|{C_1}) = S(A|{C_1}),
\end{array}
\end{equation}
since Eve ${E'}$ now is also considered to be able to control other modes ${{B_1}{C_2}{C_3}}$ and ${{C_4}}$ in the network. It can be viewed as these modes together purify the rest of modes in the system. And, for this model, only the co-variance matrix ${{\gamma _{A{C_1}}}}$ would be sufficient to calculate ${S(E')}$ and ${S(E'|{C_1})}$.

The ONUs in the network are not the only problem in the security analysis, the ODN which is located between the ONUs and the OLT is also worth of analysis.
In order to fully characterize the co-variance matrix ${{\gamma _{A{C_1}}}}$, the channel parameters, namely the transmittance and the excess noise are needed to be decided. However, since the ODN separates the entire channel into different segments, the calibration for the channel parameters should be carried out for each segment: the transmittance from the OLT to the ODN ${{T_{OLT - ODN}}}$, the transmittance from the ODN to the activated ONU ${C_1}$ ${{T_{ODN - C_1}}}$, the excess noise from the channel between the OLT to the ODN ${{\varepsilon _{OLT - ODN}}}$ and the excess noise from the channel of the ODN to the activated ONU ${C_1}$ ${{\varepsilon _{ODN - {C_1}}}}$. The ODN itself will inevitably introduce certain losses depending on split ratio of the passive BS ${{\eta _{ODN}}}$, and some possible excess noise ${{\varepsilon _{ODN}}}$. All the above parameters should be calibrated so that the elements of in co-variance matrix can be filled in. And this makes the calibration process very complicated. The solution to that is again to calibrate a total transmittance and a total excess noise since the transmittance is multiplicative and the excess noise is additive. Thus, the total transmittance and the total excess noise can be write as:
\begin{equation}
\begin{array}{l}
  {T_{tot}} = {T_{OLT - ODN}}*{T_{ODN - {C_1}}}*{\eta _{ODN}}, \hfill \\
  {\varepsilon _{tot}} = {\varepsilon _{OLT - ODN}} + {\varepsilon _{ODN - {C_1}}} + {\varepsilon _{ODN}}. \hfill \\
\end{array}
\end{equation}
In the corresponding PM model, the total transmittance and excess noise can be directly calculated from the data between the OLT and the activated ONU during the parameter estimation. It is more efficient if the channel parameters can be decided only depends on the data of the OLT and the activated ONU.

Together the upper bound information that the strengthened Eve ${E'}$ can obtain in the RR regime can then be calculated as:
\begin{equation}
{\chi _{E'{C_1}}} = S(E') - S(E'|{C_1}).
\end{equation}
Subsequently, the secret key rate can be determined by inserting ${{\chi _{E'{C_1}}}}$ into eq. (1).
Since Eve is strengthened by controlling other modes in the system, ${{\chi _{E'{C_1}}}}$ is inevitably increased compared to that of ${\chi _{E{C_1}}}$, which leads to the decrease in the obtained secret key rate compared to the standard CV-QKD. However, the security is now against all the ONUs in the network. The principle behind this is that since the standard security analysis proves the security against Eve, if Eve also includes other ONUs, the modified security analysis is secure against not only Eve, but also other ONUs in the network.
It should also be noted that the security analysis only needs to determine mode ${A}$ and mode ${C_1}$ in the system, which is exactly the two parties of the OLT and the specified ONU in the QKD, and no further calibrations are demanded. This is as well of importance in the practical implementations since the security analysis of CV-QKD between the OLT and the activated ONU should not involve other ONUs in the network.

\section{\label{sec:4} Performance analysis}

In this section, we show the performance of the downstream access network by applying the security analysis that we proved in Section~\ref{sec:3}. The one-time-calibration model~\cite{zhang2020one, huang2020modi} is applied in the secret key rate calculation, since it can model the practical imperfections of the homodyne detector and is more applicable in practical implementations, detailed secret key rate calculation is provided in the Appendix.

\begin{figure*}[htb]
\centerline{\includegraphics[width=15cm]{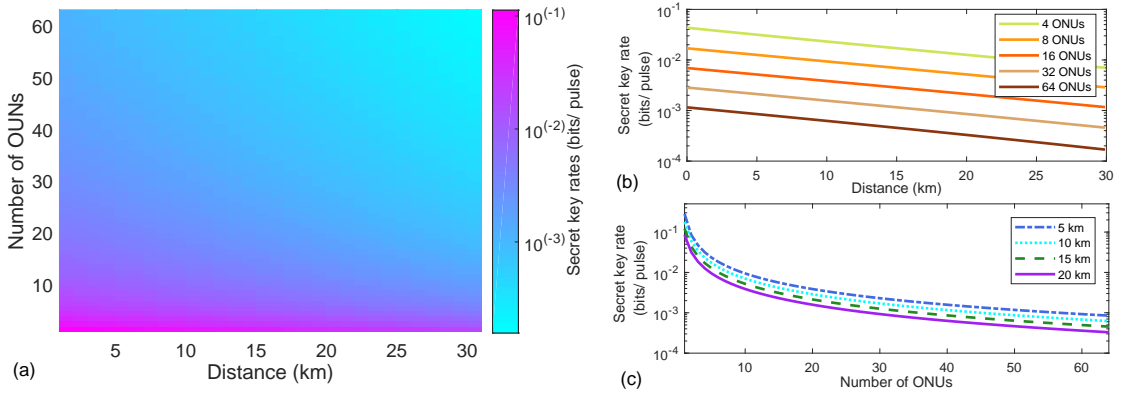}}
\caption{ (Color online) Achievable secret key rates against the number of ONUs access in the network and transmission distance. The secret key rates are represented by the pseudo-colors. Simulation parameters: variance of the EPR state ${V=5}$, channel excess noise ${\varepsilon ' = 0.05}$ (SNU), channel loss coefficient ${\alpha  = 0.2}$ dB/km, detection efficiency ${{\eta _d} = 0.6}$, electronic noise ${{\eta _e} = 0.99}$, and reconciliation efficiency ${\beta=0.956}$~\cite{zhou2019continuous, wang2018high}.
}
\end{figure*}

The achievable secret key rate which is very relevant to the number of the ONUs in the network is firstly evaluated. The transmission distance is limited at 30 km which is longer than the 20 km ITU defined maximum physical reach~\cite{itu2008recommendation}. And the number of ONUs in the access network is ranged from 2 to 64 ONUs. The simulation results of achievable secret key rate as a function of the number of ONUs and the transmission distance are shown in Fig. 3. As can be seen from Fig. 3 (a), high secret key rates can only be achieved when the number of ONUs are less than 5, when more ONUs are connecting to the network, the secret key rate drastically drops compared to the cases with less ONUs.
This is because the security is achieved not only against Eve, but also other ONUs in the network, so when more ONUs are connected to the network, the secret key rate is inevitably to be reduced.
To make the simulation clearer, secret key rate against the transmission distance alone, and against different number of ONUs alone are depicted in Fig. 3 (b) and (c). It can be seen that when more ONUs access to the network, secret key rates drops more rapidly than the increment in the total fiber distance.
Still, we observe that a positive secret key rate can be obtained even with 64 ONUs access to the network, and with transmission distance of 30 km.

The performance of the maximum tolerable excess noise after the normalization of shot-noise unit (SNU) is shown in Fig. 4, where the tolerable excess noise against different number of ONUs and the transmission distance are drawn in Fig. 4 (a). Fig. 4 (b) and Fig. 4 (c) are depicted to provide a more distinct observation on how the transmission distance and the different number of ONUs individually affect the tolerable excess noise.
It can be seen when the ONUs in the network are less than 5, the tolerable excess noise levels are significantly outperformance than other cases. Both the number of ONUs in the network and the transmission distance tend to deteriorate the maximum tolerable excess noise. And, even with 64 ONUs simultaneous access to the network, the maximum tolerable excess noise is still above zero, which reveals the feasibility for practical CV-QKD to be implemented in the downstream access network.

\begin{figure*}[htb]
\centerline{\includegraphics[width=15cm]{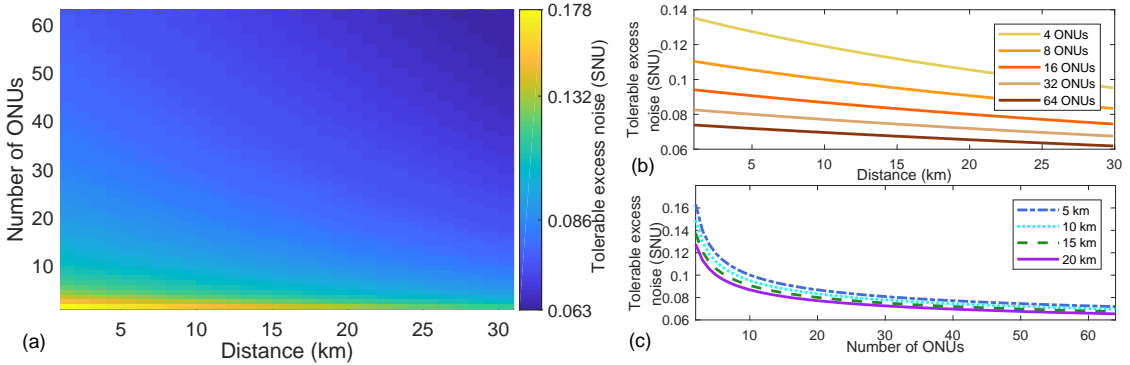}}
\caption{ (Color online) Tolerable excess noise (SNU) with different number of ONUs in the network and  different transmission distance. The pseudo-colors are used to represent the tolerable excess noise (SNU). The simulation parameters are: variance of the EPR state ${V=5}$, channel loss coefficient ${\alpha  = 0.2}$ dB/km, detection efficiency ${{\eta _d} = 0.6}$, electronic noise ${{\eta _e} = 0.99}$, and reconciliation efficiency ${\beta=0.956}$~\cite{zhou2019continuous, wang2018high}.}
\end{figure*}

In the following, the secret key rates comparison between the point-to-point CV-QKD protocol and the downstream access network protocol are demonstrated in Fig. 5. We show comparisons with different number of ONUs in the network, and with different levels of channel losses. The percentages of the secret key rate of the downstream access network compared to the standard point-to-point protocol are also shown.
It can be seen that the individual secret key rates of the downstream access network can never exceed the point-to-point scenario. Also, compare with different total fiber losses, the percentages of secret key rate ratios at low channel losses are not strictly outperformance than those at higher losses, it means that even with higher losses, the percentages of secret key rates compare to the point-to-point protocol are essentially maintained as they are in the low loss cases.

\begin{figure*}[htb]
\centerline{\includegraphics[width=17.0cm]{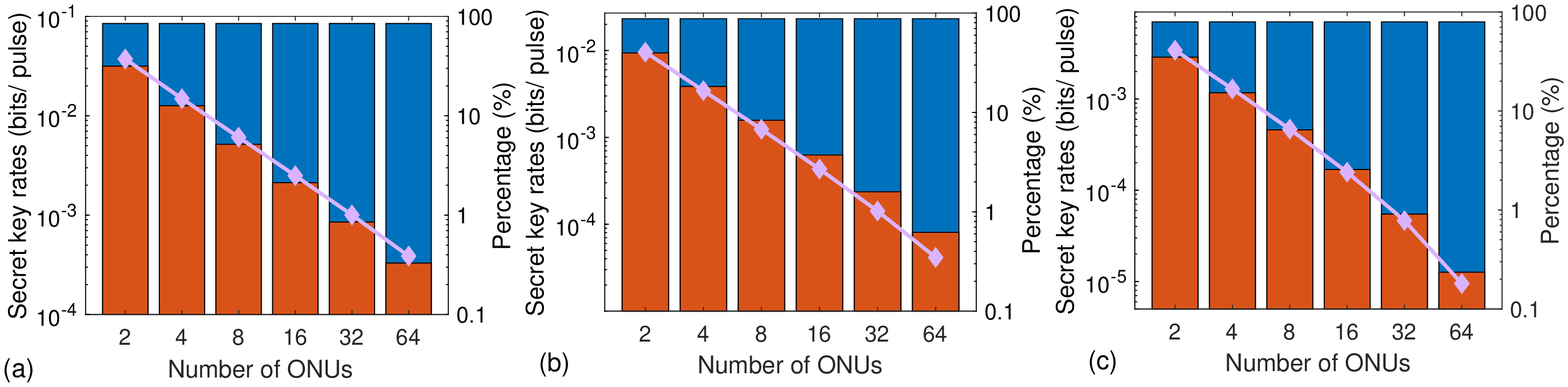}}
\caption{ (Color online) Achievable secret key rates comparisons between the downstream access network scenario and the standard point-to-point protocol. Different number of ONUs are considered in simulations. The brown bars are the secret key rates of the downstream access network, and the blue bars represent the secret key rate of the standard point-to-point scenario.
The percentages of the secret key rate of the downstream access network over the secret key rate of standard point-to-point protocol are also shown in the light violet curves. Simulations are conducted for different channel losses: (a) 4 dB, (b) 8 dB, (c) 10 dB and (d) 12 dB.
The simulation parameters are: variance of the EPR state ${V=5}$, channel excess noise ${\varepsilon ' = 0.05}$ (SNU), detection efficiency ${{\eta _d} = 0.6}$, electronic noise ${{\eta _e} = 0.99}$, and reconciliation efficiency ${\beta=0.956}$~\cite{zhou2019continuous, wang2018high}.
}
\end{figure*}

For a practical downstream access network, the OLT end can choose the optimal modulation variance for the state modulation, this flexibility may provide an improvement on the secret key rates with ONUs in the network. Subsequently, we show the optimal modulation variance with different number of ONUs and different transmission distance in Fig. 6.
It can be seen from the result, only when very limited number of ONUs accesses to the network, the optimal modulation variance drastically varies with transmission distance and the number of ONUs. When more ONUs are connecting to the network, the optimal modulation variance is rather fixed at around 4.2 (SNU), which means that the optimal modulation variance maybe picked that can help improve the secret key rates with different ONUs in the network.

\begin{figure}[t]
\centerline{\includegraphics[width=7.5cm]{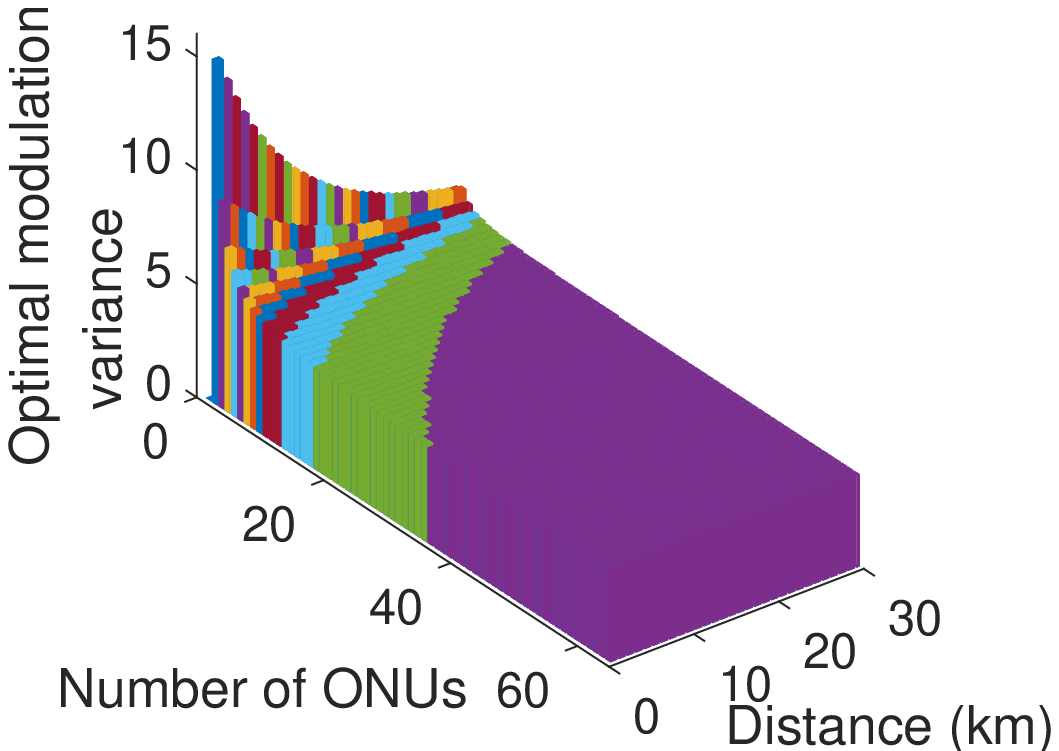}}
\caption{ (Color online) The optimal modulation variance as a function of the number of ONUs in the network, and the transmission distance. The simulation parameters are: channel excess noise ${\varepsilon ' = 0.05}$ (SNU), channel loss coefficient ${\alpha  = 0.2 }$ dB/km, detection efficiency ${{\eta _d} = 0.6}$, electronic noise ${{\eta _e} = 0.99}$, and reconciliation efficiency ${\beta=0.956}$~\cite{zhou2019continuous, wang2018high}.
}
\end{figure}

Numerous simulation results are provided in this section.
Although the secret key rate and the tolerable excess noise levels are inevitably decreased when more ONUs are connecting to the network, we still show the feasibility that up to 64 ONUs can simultaneously access to the network at the transmission distance of 30 km, which is longer than the maximum physical reach of a practical access network. Comparisons of secret key rates between the downstream access network and the standard point-to-point scenario are also demonstrated with different channel losses. Surprisingly, it occurs to us that the percentages of secret key rates of the downstream access network compare to the point-to-point protocol are essentially maintained regardless of channel loss, it means that the secret key rate performance is quite robust even against higher loss conditions. Lastly, we show the optimal modulation variance with the number of ONUs in the network as well as the transmission distance. The simulation result reveals that the optimal modulation variance may be more easily achieved when more ONUs are connecting to the network.

\section{\label{sec:5} Conclusion}
In this paper, we prove that quantum key distribution can be implemented in the downstream access network by using continuous-variable quantum key distribution.
The realization of the downstream continuous-variable quantum key distribution access network with our proposed security analysis can maximally maintain the implementation from the standard CV-QKD set-up. More specifically, both the optical line terminal Alice and the optical network units Bob can all directly inherit from the standard continuous-variable quantum key distribution implementation without much modifications. The optical distribution network only employs passive beamsplitters, and no other active control modules are needed which also greatly simplifies the control complexity.
The security analysis is shown to be secure against other optical network units in the downstream access network even though the quantum signals are broadcasted to every optical network unit.
Moreover, the security analysis is able to be conducted in a more effective way, since only data from the optical line terminal and the activated optical network unit would be sufficient to conduct the security analysis without the participation of other parties in the network.
Numerous simulation results are demonstrated to evaluate the performance of continuous-variable quantum key distribution in the downstream access network. Although the secret key rate is inevitably decreased with the increasing number of optical network units accessing to the network, the results have provided a strong feasibility of up to 64 users can be simultaneously access to the downstream access network.

For further works, the security analysis that we shown for the downstream access network can be viewed as a framework, any other protocols that are suitable for such a scenario can then be adopted into the framework.
Also, in our proposal, classical post-processing is conducted with only one activated optical network unit, meaning only one pair of symmetric secret keys are generated at a time. It is an open question of whether other optical network units in the network can use the correlation obtained from the optical line terminal to finally generate secure keys.
Lastly, other possible solutions for dealing with the security in a multiparty network scenario is considering device-independent~\cite{Vazirani2014fully} or one-sided device-independent protocols~\cite{paw2011semi, Gehring2015implementation, walk2016experimental}. These protocols generally imply that less parties are involved in the security analysis, which may become a great relaxation in the access network scenario.

Our work provides a security analysis framework for realizing quantum key distribution in the downstream access network which will help enrich the network structures for building large-scale practical quantum key distribution networks.

\begin{acknowledgments}
This work was supported by the Key Program of National Natural Science Foundation of China under Grant No. 61531003, National Natural Science Foundation of China under Grant No. 62001041, China Postdoctoral Science Foundation under Grant No. 2020TQ0016, Sichuan Science and Technology Program under Grant No. 2020YFG0289 and the Fund of State Key Laboratory of Information Photonics and Optical Communications.
\end{acknowledgments}

\appendix
\setcounter{secnumdepth}{0}
\section{Appendix: Secret key rate calculation}
The secret key rate is calculated by following the proposed security analysis for the downstream access network in Section~\ref{sec:3}, and the one-time calibration model~\cite{zhang2020one, huang2020modi} which is designed for modelling the imperfections of practical homodyne detector. Again, the coherent state and homodyne detection scheme is taken as an example in the calculation.

The security analysis of the downstream access network does not demand the participation of any other parties in the network, other than the OLT and the activated ONU. Thus, only the OLT and the activated ONU are highlighted in the EB model. The EB model in Fig. 2 (b) of the main text is an idealized model, where in practice, the homodyne detector which is deployed at the ONU suffers from imperfections. And such imperfections namely the electronic noise and the limited detection efficiency will affect the performance, thus are considered in the secret key rate calculation..

\begin{figure}[hb]
\centerline{\includegraphics[width=8cm]{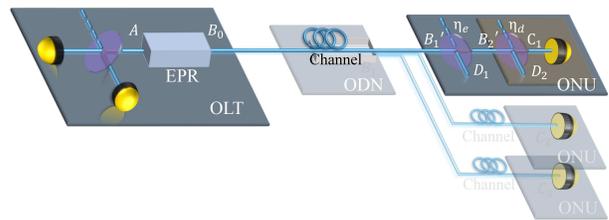}}
\caption{ (Color online) The complete EB model of the downstream access network with one-time calibration model. The coherent state and homodyne detection scheme is still applied in this model. Alice heterodyne measures one of the mode of the EPR state, and sends the other mode into the channel. When the mode arrives at the ONU, it firstly passes through the BS that imitates the electronic noise, then passes through the BS that represents the detection efficiency before being homodyne detected by the ideal homodyne detector.  }
\end{figure}

The complete EB model is shown in Fig. 7, where both the electronic noise and the limited detection efficiency are imitated by BSs. The incoming signals at the ONU first pass through the BS whose transmissivity is ${ {\eta _e}}$, then pass through the second BS with transmissivity of ${ {\eta _d}}$ before being detected by the ideal homodyne detector. Yet, for this model, only the limited detection efficiency is treated as a trusted loss, while leaving the electronic noise as part of the channel, since such a modelling can help further ease the calibration of the practical operation.

In order to calculate the secret key rate, the corresponding co-variance matrix needs to be determined. For the purpose of further obtaining the secrecy against other parties in the network, all other modes in the EB model are considered not trusted, except for the modes which are inside the OLT and the activated ONU.
Furthermore, to reduce the calibration complexity of the channel parameters, the channel parameters of different channel segments as well as the ODN are estimated as a whole, as is described in Section~\ref{sec:3}. Overall, the co-variance matrix of modes ${A, {{B'}_2}}$ and ${C_1}$ need to be determined to perform the security analysis. Suppose the channel transmissivity from the OLT to the ODN and from ODN and ONU of incoming mode ${C_1}$ are ${{T_{OLT - ODN}}}$ and ${{T_{ODN - {C_1}}}}$ respectively, and the excess noise in each channel segments are ${{\varepsilon _{OLT - ODN}}}$ and ${{\varepsilon _{ODN - {C_1}}}}$. The loss and excess noise introduced during the passive splitting are ${{T_{ODN}}}$ and ${{\varepsilon _{ODN }}}$. Instead of calibrating each parameter in the practical PM model, now we directly obtain the total channel loss including the electronic noise ${{T_{tot}} = {T_{OLT - ODN}}*{T_{ODN}}*{T_{ODN - {C_1}}}*{\eta _e}}$ and total excess noise ${{\varepsilon _{tot}} = {\varepsilon _{OLT - ODN}} + {\varepsilon _{ODN}} + {\varepsilon _{ODN + {C_1}}}}$.
Thus, the co-variance for modes ${A}$ and ${ {{B'}_2}}$ can be written as:

\begin{equation}
{\gamma _{A{{B'}_2}}} = \left( {\begin{array}{*{20}{c}}
  {V\mathbb{I}}&{\sqrt {{T_{tot}}({V^2} - 1)} {\sigma _z}} \\
  {\sqrt {{T_{tot}}({V^2} - 1)} {\sigma _z}}&{[{T_{tot}}(V - 1 + {\varepsilon _{tot}}) + 1]\mathbb{I}}
\end{array}} \right),
\end{equation}
where ${V}$ is the variance of the generated EPR state, ${{\mathbb{I}}}$ is 2*2 identity matrix and ${{{\sigma _z}}}$ is the Pauli-z matrix. Mode ${ {{B'}_2}}$ then passes through the second BS with transmissivity of ${ {\eta _d}}$, so that the co-variance matrix should be obtained as ${{\gamma _{A{C_1}{D_2}}}}$. Such a transformation can be described as:
\begin{equation}
{\gamma _{A{C_1}{D_2}}} = (\mathbb{I} \otimes {Y_{{\eta _d}}})*({\gamma _{A{{B'}_2}}} \otimes \mathbb{I})*{(\mathbb{I} \otimes {Y_{{\eta _d}}})^T},
\end{equation}
where ${{Y_{{\eta _d}}}}$ represents the transformation of the BS operation: ${{Y_{{\eta _d}}} = \left( {\begin{array}{*{20}{c}}
{\sqrt {{\eta _d}} \mathbb{I}}&{\sqrt {1 - {\eta _d}} \mathbb{I}}\\
{ - \sqrt {1 - {\eta _d}} \mathbb{I}}&{\sqrt {{\eta _d}} \mathbb{I}}
\end{array}} \right).}$

The co-variance matrix of ${{\gamma _{A{C_1}{D_2}}}}$ after the derivation should be rewritten as:
\begin{widetext}
\begin{equation}
{\gamma _{A{C_1}{D_2}}} = \left( {\begin{array}{*{20}{c}}
  {V\mathbb{I}}&{\sqrt {{T_{tot}}{\eta _d}({V^2} - 1)} {\sigma _z}}&{ - \sqrt {{T_{tot}}(1 - {\eta _d})({V^2} - 1)} {\sigma _z}} \\
  {\sqrt {{T_{tot}}{\eta _d}({V^2} - 1)} {\sigma _z}}&{[{T_{tot}}{\eta _d}(V - 1 + {\varepsilon _{tot}}) + 1]\mathbb{I}}&{ - \sqrt {(1 - {\eta _d}){\eta _d}} [{T_{tot}}(V - 1 + {\varepsilon _{tot}})]\mathbb{I}} \\
  {\sqrt {{T_{tot}}(1 - {\eta _d})({V^2} - 1)} {\sigma _z}}&{ - \sqrt {(1 - {\eta _d}){\eta _d}} [{T_{tot}}(V - 1 + {\varepsilon _{tot}})]\mathbb{I}}&{[{T_{tot}}(1 - {\eta _d})(V - 1 + {\varepsilon _{tot}}) + 1]\mathbb{I}}
\end{array}} \right).
\end{equation}
\end{widetext}

After the rearrangement, the final co-variance matrix ${{\gamma _{A{D_2}{C_1}}}}$ can be figured out.
The final secret key rate of the security analysis is calculated as:
\begin{equation}
K = \beta {I_{A{C_1}}} - {\chi _{E'{C_1}}},
\end{equation}
where ${\beta }$ is the reconciliation efficiency and ${{I_{AC}}}$ and ${{\chi _{E'{C_1}}}}$ are the mutual information between the OLT Alice and the specified ONU, and the information that Eve can obtain during the reverse reconciliation, and each terms can then be determined from the corresponding co-variance matrix ${{\gamma _{A{D_2}{C_1}}}}$.
The first term of eq. (9) ${{I_{A{C_1}}}}$ is calculated as:
\begin{equation}
{I_{A{C_1}}} = \frac{1}{2}\log \frac{{{V_A} + 1}}{{{V_{A|{C_1}}} + 1}},
\end{equation}
where ${V_A}$ is the variance of mode ${A}$. ${{{V_{A|{C_1}}}}}$ is the conditional variance of mode ${A}$ given the measurement result of mode ${C_1}$, which can be calculated as: ${{V_{A|{C_1}}} = {V_A} - \frac{{ < A{C_1} > }}{{{V_{{C_1}}}}}}$, where ${{ < A{C_1} > }}$ is the co-variance of mode ${A}$ and mode ${C_1}$ and ${{{V_{{C_1}}}}}$ is the variance of mode ${C_1}$, they can all be determined from the co-variance matrix ${{\gamma _{A{D_2}{C_1}}}}$.
${{\chi _{EC}}}$ is composed of two parts:
\begin{equation}
{\chi _{E'{C_1}}} = S(E') - S(E'|x_{{C_1}}^M).
\end{equation}
The first part ${S(E')}$ equals to ${S(A{D_2}{C_1})}$ as is described in Section ~\ref{sec:3} of the main text. ${ S(E'|x_{{C_1}}^M)}$ represents the Von Neumann entropy that after the mode ${C_1}$ is being homodyne detected. Again, Eve can purify the rest of the mode in the system as ${ S(E'|x_{{C_1}}^M)}$ equals to ${ S(A{D_2}|x_{{C_1}}^M)}$, which can be determined from the co-variance matrix ${ \gamma _{A{D_2}}^{{m_{{C_1}}}}}$. The transformation can be described as:
\begin{equation}
\gamma _{A{D_2}}^{{m_{{C_1}}}} = {\gamma _{A{D_2}}} - {\sigma _{A{D_2}}}{(X{\gamma _{{C_1}}}X)^{MP}}\sigma _{A{D_2}}^T,
\end{equation}
where ${X = diag(1,\;0,\;0,\;0)}$ and ${MP}$ stands for the inverse on the range.
${ {\gamma _{A{D_2}}}}$, ${{{\sigma _{A{D_2}}}}}$ and ${{{\gamma _{{C_1}}}}}$ can all be find out in the co-variance matrix ${{\gamma _{A{D_2}{C_1}}} = \left( {\begin{array}{*{20}{c}}
{{\gamma _{A{D_2}}}}&{{\sigma _{A{D_2}}}}\\
{\sigma _{A{D_2}}^T}&{{\gamma _{{C_1}}}}
\end{array}} \right)}$.
By inserting these terms back to eq. (13), the secret key rate can be calculated.


\end{document}